\documentclass[preprintnumbers,aps,prl,preprint,showpacs,showkeys]{revtex4}

\usepackage{wrapfig}
\usepackage{graphicx}
\usepackage{amsmath}
\usepackage{amssymb}
\usepackage{amsfonts}

\begin{document}

\preprint{APS/123-QED}

\title{Understanding the atomic-scale contrast in Kelvin Probe Force Microscopy}

\author{Laurent Nony$^{1, 2, }$\footnote{Corresponding author:
laurent.nony@im2np.fr.}, Adam S. Foster$^{3,4}$, Franck
Bocquet$^{1, 2}$ and Christian Loppacher$^{1, 2}$}
\affiliation{$^{(1)}$Aix-Marseille Universit\'{e}, IM2NP, Av.
Normandie-Niemen, Case 151, F-13397 Marseille CEDEX 20, France and
$^{(2)}$CNRS, IM2NP (UMR 6242), Marseille-Toulon, France\\
$^{(3)}$Department of Physics, Tampere University of Technology
P.O. Box 692, FIN-33101 Tampere, Finland\\
$^{(4)}$Department of Applied Physics, Helsinki University of Technology,
P.O. Box 1100, FIN-02015 HUT, Finland}

\date{\today}



\begin{abstract}
A numerical analysis of the origin of the atomic-scale contrast in
Kelvin probe force microscopy (KPFM) is presented. Atomistic
simulations of the tip-sample interaction force field have been
combined with a non-contact Atomic Force Microscope/KPFM
simulator. The implementation mimics recent experimental results
on the (001) surface of a bulk alkali halide crystal for which
simultaneous atomic-scale topographical and Contact Potential
Difference (CPD) contrasts were reported. The local CPD does
reflect the periodicity of the ionic crystal, but not the
magnitude of its Madelung surface potential. The imaging mechanism
relies on the induced polarization of the ions at the tip-surface
interface owing to the modulation of the applied bias voltage. Our
findings are in excellent agreement with previous theoretical expectations
and experimental observations.

Published in Phys. Rev. Lett. \textbf{103}, 036802 (2009)
\end{abstract}

\maketitle

Kelvin Probe Force Microscopy (KPFM) is a scanning probe technique
able to compensate dynamically for electrostatic forces occurring
between biased tip and sample \cite{weaver91a}. These forces stem
from work function differences between the electrodes when they do
not consist of similar materials and/or when they carry charges.
The KPFM controller supplies the proper DC voltage to align the
Fermi levels of both electrodes, thereby compensating the
electrostatic force. Thus, KPFM provides access to the contact
potential difference (CPD) between electrodes. On the experimental
level, KPFM is combined with the noncontact-Atomic Force
Microscopy operating mode (nc-AFM) \cite{kitamura98a}. Hence,
topographical and CPD images are acquired simultaneously.
\newline \indent For a decade, KPFM has proven its ability
to map CPD changes at the nanometer scale with an accuracy of a
few millivolts \cite{rosenwaks04a}. On various semi-conducting
surfaces, several groups reported CPD images with contrasts
showing atomic features, the positions of which were consistent
with the ones of the surface atoms \cite{sugawara99a, kitamura00a,
okamoto03a, krok08a}. However, some of these early works pointed
out the inconsistency between the measured CPD and the values
derived from other experimental techniques or from theoretical
calculations \cite{kitamura98a,kitamura00a,okamoto03a}. It was
stated that, when measured close to the surface, the so-called
``local CPD" (LCPD) is influenced by the surface potential. Thus,
it differs from the intrinsic CPD that relies on the macroscopic
concept of work function \cite{wandelt97a}. The physics that
connects the magnitude of the LCPD and the surface potential still
remains under debate. Recently, Loppacher \emph{et al.} have
described a self-consistent analytical approach to the LCPD probed
by KPFM with a metallic tip on the (001) facet of a bulk KBr
crystal \cite{bocquet08a,nony09a}. It was found that the LCPD
neither reflects the actual tip-surface CPD, nor the local surface
potential, but an effective potential following the atomic
corrugation of the Madelung surface potential of the ionic crystal
\cite{watson81a}, but differing from it due to the strong
influence of the geometry of the tip. The theoretical framework
proposed so far helps in understanding the origin of the
atomic-scale KPFM contrast, but relies on an unrealistic geometry
of the metallic tip. Thus, the analytical approach remains
questionable, as it is known that the atomic-scale topographical
contrast on alkali halides stems from tips apexes carrying ionic
clusters, preliminarily picked-up from the surface
\cite{hoffmann04a}.
\newline \indent In this letter, we conjointly improve the description
of the tip and its interaction with the sample in order to achieve
simultaneous atomic-scale topographical and KPFM contrast on a
(001) NaCl surface. The work is carried out on the base of
atomistic simulations of the force field occurring between a
realistic tip consisting of a metallic body carrying a NaCl
cluster and a defect-free (001) NaCl surface. Topographical and
LCPD images are computed by means of our non-contact AFM/KPFM
simulator \cite{nony06a,nony09a} operated in the frequency
modulation-KPFM mode (FM-KPFM). For the first time, particular
attention is paid to the ionic polarization at the tip-surface
interface, i.e. the displacements of the ions induced by the
inherent modulation of the applied bias voltage in the KPFM
technique. Although weak (i.e. few picometers), and hence
neglected so far, these displacements are crucial to understand
the atomic-scale KPFM signal.

\begin{figure}
    \centering
        \includegraphics[width=\columnwidth]{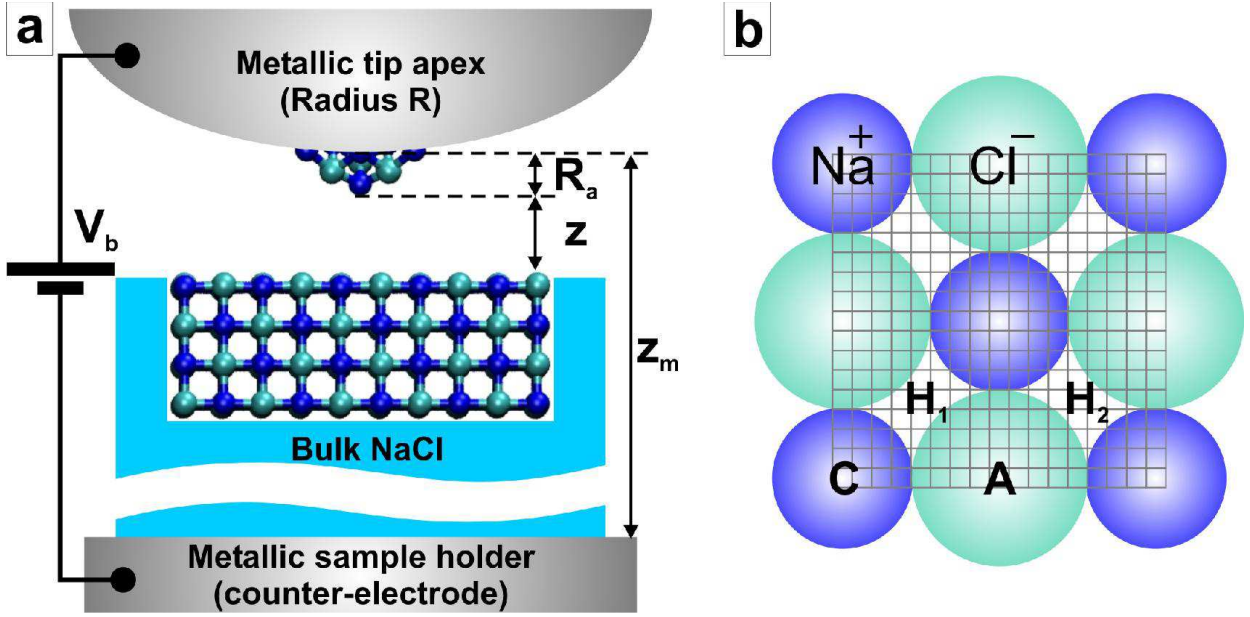}
    \caption{(Color online) a- Sketch of the numerical tip-surface setup. We have set $z_m=5$~mm compared to $z$ which scales in the sub-nm range.
    b- Sketch of the NaCl unit cell showing the 17$\times$17 mesh used to calculate the (x,y,z,V) four-dimensional tip-surface force field. We have focused on four particular sites: anionic ($A$), cationic ($C$) and hollow ($H_1$, $H_2$) sites.}
    \label{FIG_GEO}
\end{figure}

The calculations of the force field were performed using atomistic
simulations as implemented in the code
SCIFI~\cite{kantorovich00a}. The interatomic forces are computed
from a sum of pairwise Buckingham potentials acting between ions
treated atomistically in a shell model with coupled oppositely
charged cores and shells in order to describe their
polarizabilities. The SCIFI code also allows for the inclusion of
metallic electrodes at the tip and below the surface. The
interaction of these with ions in the surface and tip are treated
by the method of images \cite{kantorovich00a}. Using this
approach, we can simulate the polarization of conductors and
resultant atomic geometries in the system as a function of tip
position and applied bias voltage. Parameters for the species
considered were taken from Ref.\cite{shluger94a}. Unless specified
otherwise, all cores and shells were allowed to relax completely
with respect to interatomic and image forces with a convergence
criterion of 1~meV/\AA~per ion. Hence, in this work we take into
account ionic relaxation, and electronic and ionic polarization as
a function of both atomic interactions and applied bias.
\newline \indent The properties of the NaCl(001) surface are well understood and
can be well represented by a slab of 4 atomic layers containing
$10 \times 10$ ions, with those in the bottom layer and edges kept
fixed (cf. Fig.\ref{FIG_GEO}a). The NaCl slab is embedded within a
semi-infinite, 5~mm thick, slab merely treated by means of its
dielectric constant. A 64-atom cubic cluster of NaCl is embedded
into a metallic sphere of radius $R=$5~nm and oriented such that
the [111] direction is perpendicular to the surface with a Na atom
at the apex. The most stable configuration is found when the
cluster protrudes from the end of the sphere with a height
$R_A=0.3$~nm (cf. Fig.\ref{FIG_GEO}a), and carries an intrinsic
charge of +1 inducing an opposite charge in the vicinity of the
metallic part of the tip. The NaCl atoms within the sphere are
frozen and play no role in the calculation of image forces. They
act as ghost metal atoms stabilizing the tip apex. The metallic
part of the tip is biased with respect to the counter-electrode
holding the crystal.
\newline \indent In order to compute images with the simulator, the NaCl unit cell
was meshed with a 17$\times$17 grid (cf. Fig.\ref{FIG_GEO}b). We
have focused at four particular sites: anionic site $A$, cationic
site $C$ and hollow sites $H_1$ and $H_2$ which are made
inequivalent owing to the orientation of the cluster with respect
to the surface symmetry. For each pixel of the mesh, the distance
dependence ($z$-dependence) of the force is computed by 5~pm steps
from 0.3~nm to 2.0~nm and the bias dependence by 10~mV steps from
-3.4~V to +2.3~V. A Van der Waals force has been added as a
long-range background (equ.2.4 in Ref.\cite{guggisberg00a}). A
long-range electrostatic force has been included as well with the
form given in Ref.\cite{nony09a} (equ.14). It stands for a
phenomenological term which depicts capacitive effects between the
cantilever and the counter-electrode when mounted in the
microscope. When using the simulator, we have made sure that the
controllers are all in a critically damped regime when the tip is
close to the surface, which prevents imaging artifacts from
occurring \cite{nony06a}. The main parameters of the simulations
are: oscillation amplitude: 8~nm peak-to-peak, resonance
frequency: 150~kHz, cantilever stiffness: 30~N/m, $Q$-factor:
30000, scan size: 1.03$\times$1.03~nm$^2$, scan speed: 1.5~s/line.
The intrinsic charge of the tip induces a long-range background
force. Thus, a DC voltage, $V_{dc}^\text{ref}=-0.91$~V, is applied
to the tip to compensate for it and nullify the LCPD at large
tip-sample separation (i.e. $>$2~nm). $V_{dc}^\text{ref}$ can be
interpreted as the opposite of the macroscopic CPD of the
electrodes-bulk NaCl system. Thus, the bias voltage applied to the
tip is $V_b=V_{dc}+V_{dc}^\text{ref}+V_{ac}\sin(2\pi
f_\text{mod}t)$. The FM-KPFM mode was implemented with a 500~Hz
bandwidth lockin amplifier and a 50~Hz bandwidth controller. The
AC bias modulation is $V_{ac}=$0.5~V and $f_\text{mod}=1$~kHz.

\begin{figure}[t]
    \centering
        \includegraphics[width=\columnwidth]{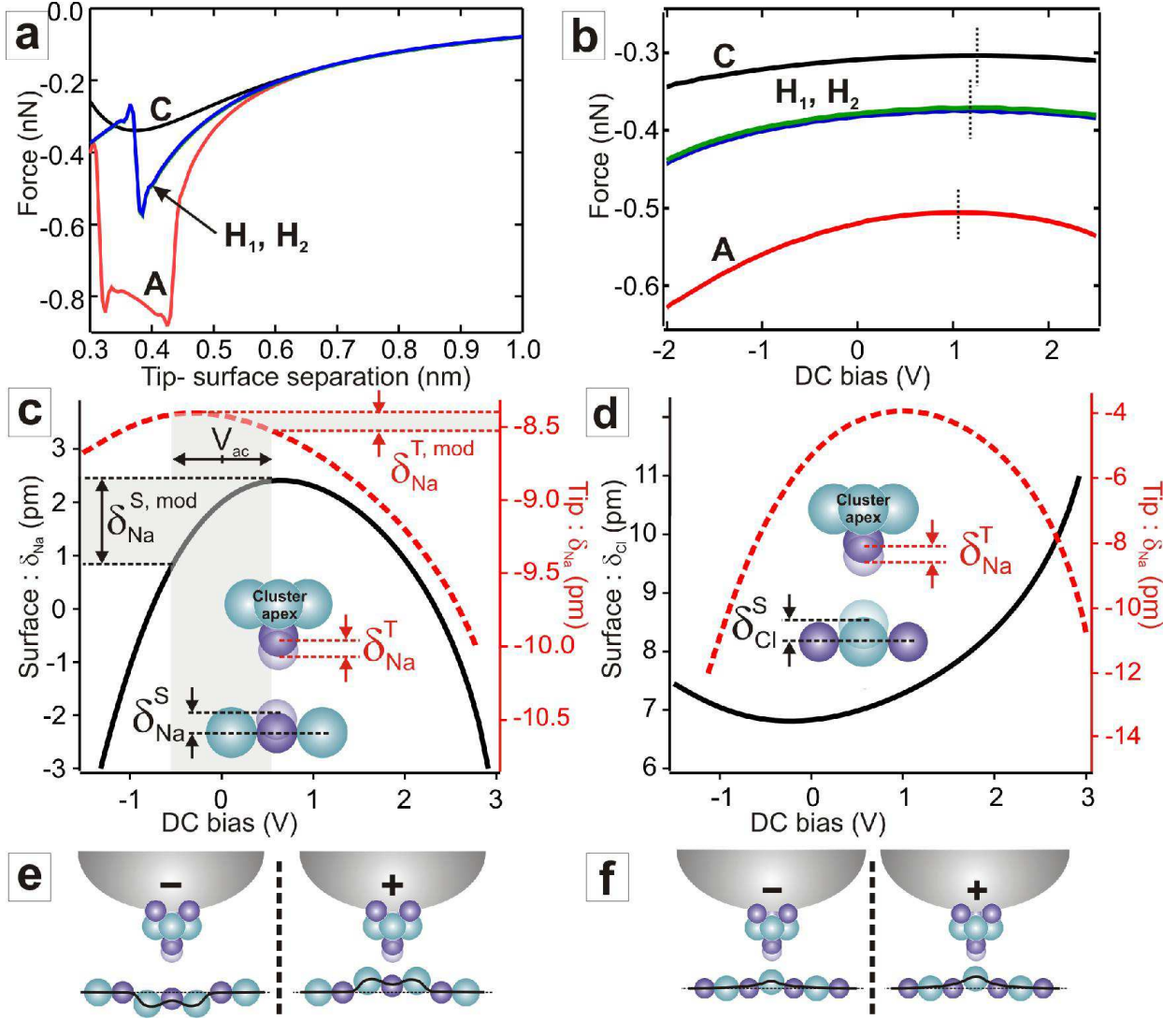}
    \caption{(Color online) a- Force vs.~distance curves measured above the four sites at $V_{dc}=0$~V. Below 0.45~nm the tip becomes unstable. b- Force vs.~$V_{dc}$ curves at $z=0.45$~nm. The dependence is not parabolic.
    c- $V_{dc}$ dependence of the displacement of the foremost Na$^+$ ion of the tip, $\delta_\text{Na}^\text{T}$ (dashed line), at $z=0.45$~nm on top of Na$^+$ of the slab and corresponding $\delta_\text{Na}^\text{S}$ displacement.
    The AC bias modulation (grey) triggers the dynamic displacement of the ions at the interface ($\delta_\text{Na}^\text{T, S, mod}$). d- Same as c- except that the tip is now placed on top of Cl$^-$.
    Scheme of the induced ionic displacements as a function of the sign of the bias voltage on top of e- Na$^+$, and f- Cl$^-$.}
    \label{FIG_FORCE}
\end{figure}

Force vs.~distance curves computed above the four sites with
$V_{dc}=0$~V are shown in Fig.\ref{FIG_FORCE}a. Below 0.45~nm,
tip/surface instabilities on top of anionic and hollow sites
occur. Above 0.45~nm, the curves differ significantly, although
exhibiting similar features to those reported with almost
equivalent setups \cite{hoffmann04a,lantz06a,schirmeisen06a}.
Force vs.~$V_{dc}$ curves measured at $z=0.45$~nm are shown in
Figs.\ref{FIG_FORCE}b for the four sites. The maxima of the curves
differ between sites (cf. dotted lines): -304~pN at 1.22~V (site
$C$) and -506~pN at 1.06~V (site $A$). The curves systematically
deviate from the capacitive, parabolic-like, behavior which stems
from the polarization of the ions at the tip-surface interface. To
assess this, the displacements of the cores of the foremost Na$^+$
ion of the tip ($\delta_\text{Na}^\text{T}$) when placed above a
Na$^+$ ($\delta_\text{Na}^\text{S}$) and above a Cl$^-$
($\delta_\text{Cl}^\text{S}$) of the slab as a function of
$V_{dc}$ are shown in Figs.\ref{FIG_FORCE}c and d, respectively.
They are measured at $z=0.45$~nm. A positive displacement means
that the ion is displaced upwards (e.g. towards the tip when
considering an ion of the slab). We only have focused on the
displacements of the cores of the ions that were judged as the
most significant, although the polarization process involves all
the ions of the interface and their shells. For the sake of
clarity, we have as well sketched the ionic displacements in
Figs.\ref{FIG_FORCE}e and f.

On top of Na$^+$ at zero bias, the foremost cation of the tip is
attracted towards the surface:
$\delta_\text{Na}^\text{T}$=-8.5~pm. Simultaneously, the Na$^+$ of
the slab undergoes a moderate displacement towards the tip:
$\delta_\text{Na}^\text{S}$=+2~pm. This behavior stems from the
balance between the short-range chemical interaction and the local
electrostatic interaction due to the intrinsic charge of the tip,
merely compensated by $V_{dc}^\text{ref}$ at large distance. With
$V_{dc}>0$, the foremost cation of the tip remains attracted to
the surface, while the Na$^+$ is repelled within the slab. The
short-range electrostatic force is then strengthened between the
tip cation and the four Cl$^-$ closest neighbors of the Na$^+$ of
the slab, while the latter is repelled from the tip because of the
overall less favourable chemical and electrostatic interaction.
With $V_{dc}<0$, the electrostatic force becomes dominant and
mostly repulsive for the same reason as before. Then, the set of
Cl$^-$ and Na$^+$ ions are repelled within the slab.

On top of Cl$^-$ at zero bias, the favourable combination of the
chemical interaction and of the local electrostatic interaction
due to the intrinsic charge of the tip partly compensated by
$V_{dc}^\text{ref}$ produces significant displacements of the ions
at the interface ($\delta_\text{Na}^\text{T}=-6$~pm;
$\delta_\text{Cl}^\text{S}=+7$~pm). With $V_{dc}>0$, the local
electrostatic interaction increases the mutual attraction between
ions. With $V_{dc}<0$, the Cl$^-$ of the slab is less attracted by
the tip due to the repulsive electrostatic interaction, but the
tip cation remains attracted by the surface, likely because the
chemical interaction is still large enough.

Hence, when the KPFM controller is engaged, the AC modulation of
the bias triggers complex dynamic displacements of the
cluster/surface ions. As predicted in Ref.\cite{bocquet08a}, these
displacements support the LCPD signal and explain the deviation
from the usual capacitive parabolic law of the force vs.~bias
voltage curve. Indeed, when performing the following simulations
while freezing the ionic polarization, no KPFM contrast occurs.

\begin{figure}[t]
    \centering
        \includegraphics[width=\columnwidth]{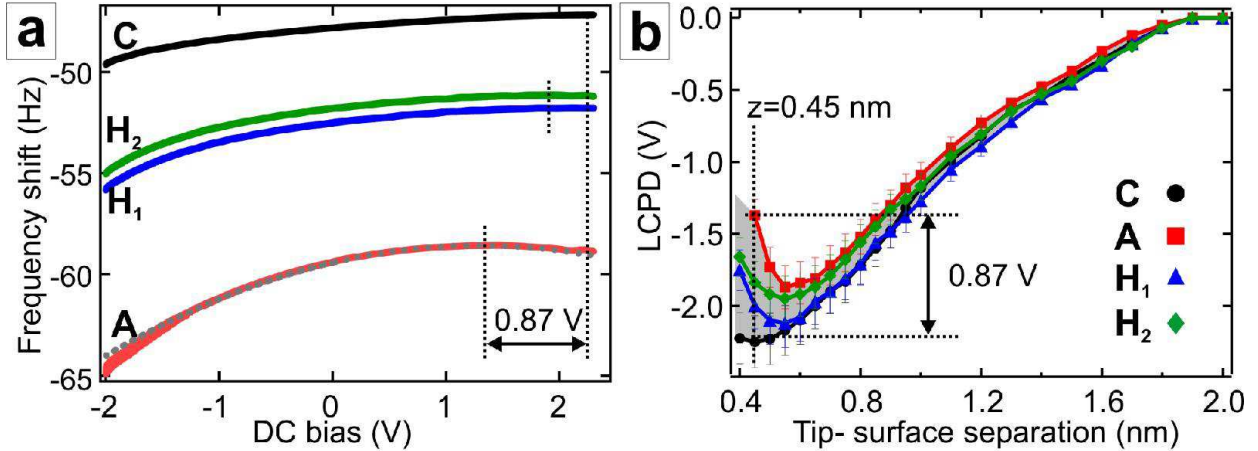}
    \caption{(Color online) a- Spectroscopic curves computed above the four sites at $z=0.45$~nm. A shift of 0.87~V is noticed between anionic ($A$) and cationic ($C$) sites.
    b- Distance dependence of the LCPD above the four sites derived from the spectroscopic curves. In the short-range regime, the LCPD exhibits a resonance-like and site-dependent behavior.}
    \label{FIG_LCPD}
\end{figure}

The distance dependence of the LCPD has been investigated by means
of spectroscopic curves, during which the KPFM and distance
controllers are disengaged. When the tip is biased, the maximum of
the curve gives a DC voltage opposite to that of the LCPD:
$V_{dc}=-V_\text{LCPD}$. In Fig.\ref{FIG_LCPD}a are shown
spectroscopic curves measured on top of each site at $z=0.45$~nm.
As expected from the force vs.~$V_{dc}$ curves, the spectroscopic
curves deviate from the parabolic-like behavior (shown for site
$A$, dotted greyed curve) and the positions of the maxima differ
upon sites. Furthermore, the latter positions do not match those
of the force vs.~$V_{dc}$ curves. However, such an effect is
expected to occur as soon as the $z$ and $V$ dependencies of the
interaction force cannot be separated, i.e. $F(z,V_{dc}) \neq
h(z)g(V_{dc})$. A shift of +0.87~V is measured from site $A$ to
site $C$, consistently with the larger repulsive electrostatic
force observed above cations. These measurements have been
reproduced for various tip-surface separations and gathered in
Fig.\ref{FIG_LCPD}b. When increasing the separation, the LCPD
first decreases and then increases to converge towards 0 at large
distance, as stated before. Below 0.6~nm, the curves unbundle and
differ significantly upon sites (greyed area). These curves are
equivalent to $\Delta f$ vs.~distance curves that are driving the
magnitude of the topographical contrast. Hence, a site-dependent
KPFM contrast is indeed expected while scanning for tip-surface
separations smaller than 0.6~nm. The magnitude of the LCPD
contrast can be derived as well. At $z=0.45$~nm (dotted line), a
maximum of 0.87~V is expected. At equivalent height, the Madelung
surface potential is 0.14~V \cite{bocquet08a}. This resonance-like
effect has been predicted theoretically \cite{nony09a} and
reported experimentally \cite{burke09a}. It relies on a subtle
balance between short-range and long-range electrostatic forces,
both weighting in the manner the LCPD is compensated.

\begin{figure}[t]
    \centering
        \includegraphics[width=\columnwidth]{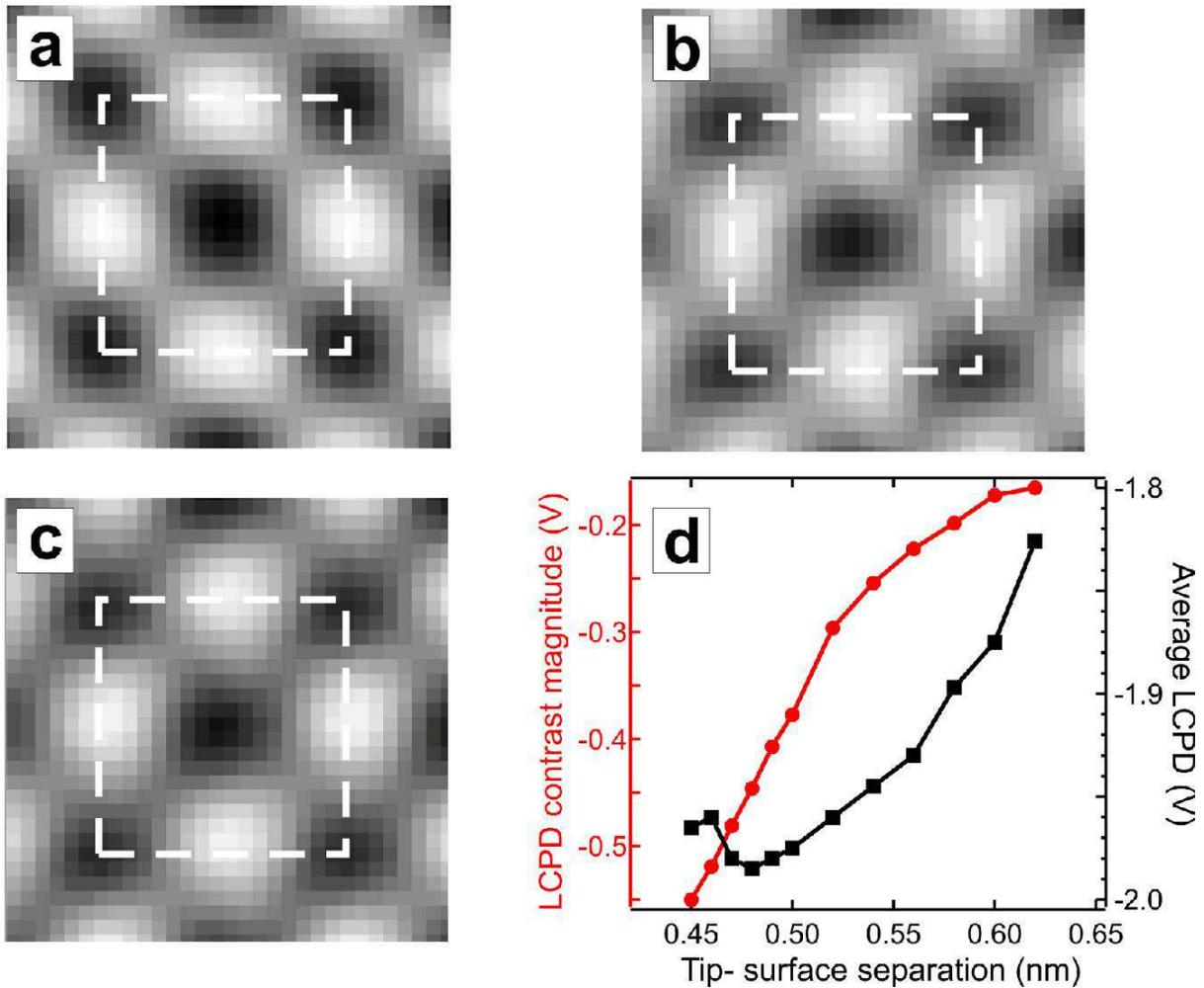}
    \caption{(Color online) a- Topographical image computed with the nc-AFM/KPFM simulator. The vertical contrast is 38~pm. b- Simultaneously computed LCPD image. The contrast ranges from -2.24 to -1.69~V (0.56~V full scale).
    c- LCPD image computed at constant height, $z=0.45$~nm. The contrast ranges from -2.24 to -1.38~V (0.86~V full scale), consistently with the expected range deduced from Fig.\ref{FIG_LCPD}b.
    d- Evolution of the magnitude of the LCPD contrast (dots) and of the average LCPD (squares) as a function of the distance.}
    \label{FIG_IMAGES}
\end{figure}

Finally, topographical and LCPD images have been computed (cf.
Figs.\ref{FIG_IMAGES}a and b, c, respectively). Images shown in
Figs.\ref{FIG_IMAGES}a (38~pm full scale) and b (0.56~V full
scale) have been simultaneously computed with the distance
controller engaged. The scan has been engaged on top of a cation
at $z=0.45$~nm, corresponding to $\Delta f_{set}=-47.22$~Hz. The
dotted area depicts the unit cell shown in Fig.\ref{FIG_GEO}b.
Topographical and LCPD images cations as depressions, consistently
with spectroscopic curves. The magnitude of the contrasts as well
as the distance range are in good agreement with our former
experimental observations (30~pm, 0.1~V) \cite{bocquet08a}.
Fig.\ref{FIG_IMAGES}c (0.86~V full scale) is a LCPD image computed
with similar conditions as b, but at constant height $z=0.45$~nm,
i.e. with the distance controller disengaged. The magnitude of the
contrast matches the predicted behavior (cf. Fig.\ref{FIG_LCPD}b,
dotted line). In Fig.\ref{FIG_IMAGES}d is reported the magnitude
of the LCPD contrast as a function of the tip-surface separation.
The curve is deduced from scans for which the distance regulator
was engaged. The average value of the LCPD (mean contrast) has
been reported as well. It follows accurately the evolution of the
average LCPD derived from the spectroscopic curves (cf.
Fig.\ref{FIG_LCPD}b). The contrast expands around the average
value while keeping confined within the greyed area, the size of
which is controlled by the combination between short-range
electrostatic and chemical forces. Thus, relevant information
about the LCPD is not only carried by the magnitude of the KPFM
contrast, but also by its average value.

In conclusion, we have described an imaging mechanism which
accounts for simultaneous occurrence of atomic-scale topographical
and KPFM contrast on a perfect bulk alkali halide crystal. A
realistic tip design and a fine description of the bias and
distance dependence of the interaction force field were mandatory
to explain the KPFM contrast. Our major finding is that the
modulation of the bias voltage triggers short-range electrostatic
forces which induce dynamic displacements of the ions at the
tip-surface interface. These render the short-range chemical
forces self-consistently bias-dependent. As the LCPD stands for
the DC bias that makes the total force maximum, its mere
interpretation on the basis of the Madelung surface potential is
insufficient. Nevertheless, our results obtained on a defect-free
ionic surface are crucial in order to quantitatively interpret
KPFM measurements on more complex surfaces including defects.

ASF wishes to thank L. N. Kantorovich for useful discussions and
acknowledges support from the Academy of Finland and ESF FANAS
programme.


\begin{thebibliography}{10}

\bibitem{weaver91a}
J. Weaver and D. Abraham, J. Vac. Sci. Technol. B {\bf 9},  1559
(1991).

\bibitem{kitamura98a}
S. Kitamura and M. Iwatsuki, Appl. Phys. Lett. {\bf 72},  3154
(1998).

\bibitem{rosenwaks04a}
Y. Rosenwaks, R. Shikler, T. Glatzel, and S. Sadewasser, Phys.
Rev. B {\bf 70},
   085320  (2004).

\bibitem{sugawara99a}
Y. Sugawara, T. Uchihashi, M. Abe, and S. Morita, Appl. Surf. Sci.
{\bf 140},
  371  (1999).

\bibitem{kitamura00a}
S. Kitamura, K. Suzuki, M. Iwatsuki, and C. Mooney, Appl. Surf.
Sci. {\bf 157},
   222  (2000).

\bibitem{okamoto03a}
K. Okamoto, K. Yoshimoto, Y. Sugawara, and S. Morita, Appl. Surf.
Sci. {\bf
  210},  128  (2003).

\bibitem{krok08a}
F. Krok {\it et~al.}, Phys. Rev. B {\bf 77},  235427  (2008).

\bibitem{wandelt97a}
K. Wandelt, Appl. Surf. Sci. {\bf 111},  1  (1997).

\bibitem{bocquet08a}
F. Bocquet, L. Nony, C. Loppacher, and T. Glatzel, Phys. Rev. B
{\bf 78},
  035410  (2008).

\bibitem{nony09a}
L. Nony, F. Bocquet, C. Loppacher, and T. Glatzel, Nanotechnology
{\bf 20}, 264014 (2009).

\bibitem{watson81a}
R. Watson, J. Davenport, M. Perlam, and T. Sham., Phys. Rev. B
{\bf 24},  1791
  (1981).

\bibitem{hoffmann04a}
R. Hoffmann {\it et~al.}, Phys. Rev. Lett. {\bf 92},  146103
(2004).

\bibitem{nony06a}
L. Nony {\it et~al.}, Phys. Rev. B {\bf 74},  235439  (2006).

\bibitem{kantorovich00a}
L. Kantorovich, A. Foster, A. Shluger, and A. Stoneham, Surf. Sci.
{\bf 445},
  283  (2000).

\bibitem{shluger94a}
A. Shluger, A. Rohl, D. Gay, and R. Williams, J. Phys.: Condens.
Matter {\bf
  6},  1825  (1994).

\bibitem{guggisberg00a}
M. Guggisberg {\it et~al.}, Phys. Rev. B {\bf 61},  11151  (2000).

\bibitem{lantz06a}
M. Lantz {\it et~al.}, Phys. Rev. B {\bf 74},  245426  (2006).

\bibitem{schirmeisen06a}
A. Schirmeisen, D. Weiner, and H. Fuchs, Phys. Rev. Lett. {\bf
97},  136101 (2006).

\bibitem{burke09a}
S. Burke {\it et~al.}, Nanotechnology {\bf 20}, 264012 (2009).

\end{thebibliography}
\end{document}